# Pursuing More Sustainable Energy Consumption by Analyzing Sectoral Direct and Indirect Energy Use in Malaysia: An Input-Output Analysis

HARUN MUKARAMAH[a*], HASSAN SALLAHUDDIN[a] AND SITI HADIJAH CHE MAT[b]

[a]*School of Economic, Finance and Banking, Universiti Utara Malaysia, Malaysia*
[b]*College of Business, Universiti Utara Malaysia, Malaysia*

## ABSTRACT

Malaysia is experiencing ever increasing domestic energy consumption. This study is an attempt at analyzing the changes in sectoral energy intensities in Malaysia for the period 1995 to 2011. The study quantifies the sectoral total, direct, and indirect energy intensities to track the sectors that are responsible for the increasing energy consumption. The energy input-output model which is a frontier method for examining resource embodiments in goods and services on a sectoral scale that is popular among scholars has been applied in this study. The results reveal that both the direct and indirect sectoral energy intensities are important contributors to total sectoral energy intensity. Therefore, for effective efficiency policy, different strategies should be adopted based on a particular sector's direct and indirect energy intensity features. The sectors with the highest potentials for conservation, which are also the main consumers of energy and have the highest energy intensities are transport and storage, non-metallic mineral products, chemicals and chemical products, financial intermediation, agriculture, forestry, and fishing, fabricated metal products, and food, beverage and tobacco. By investigating both the direct and the indirect energy usage for sectors in Malaysia, this study adds to researches focusing only on the direct sectoral energy requirement in Malaysia.

* Corresponding author: Email: mukaramah@uum.edu.my

## INTRODUCTION

Over the last two decades, Malaysia has experienced a substantial increase in energy consumption. In 1995, total energy demand[1] was 23,101 Ktoe, but by 2011, it had reached 72,737 Ktoe. This translates to the Malaysian economy experiencing, on average 7.4 percent increase in total energy demand per year over the last sixteen years. Expansion of economic activities especially in the manufacturing, transportation and residential sectors, urbanization and growing population (increased from 15.9 million in 1995 to 29.1 million in 2011) have been the major drivers for the increasing demand for the energy (Malaysia National Energy Master Plan, 2014; Razali et al., 2015; Shahbaz et al., 2015). The increase in energy consumption has also been driven by the Malaysian fuel price control policy[2].

Historically, for Malaysia, total energy demand growth rates were higher than the growth rates of GDP. For the period 1995 to 2011, the growth in total energy demand was 9.6 percent per year, while the growth in GDP was 6.3 percent per year.[3] The imbalance between the growth rates of energy demand and GDP is indicative of the more energy-intensive economic activities driving the growth. As Malaysia is determined to maintain its economic growth over the next decades, and realizing that the growth in its energy consumption must be managed to ensure the productivity and competitiveness of its economy, the Malaysian government has launched several programs to promote energy efficiency. The well-known Malaysia Efficiency Master Plan was formulated in 2010. This was in tandem with the national Tenth Malaysia Plan (2011-2015) which aims for energy efficiency improvement as one of the important agenda in the plan. The Malaysia Efficiency Master Plan (2014)[4] presents a strategy for a well-coordinated and cost-effective implementation of energy efficiency measures in sectoral activities which will lead to lower energy consumption in the country.

Motivated by the government's efforts, this study intends to provide insights on the energy efficiency performance of sectoral activities and how they are interconnected. Energy efficiency is the output (often in non-energy units) per unit input of energy. The inverse of energy efficiency is called "energy intensity". Increasing energy efficiency (or equivalently: reducing energy intensity), can help reduce energy consumption and thus conserve energy. Energy intensity is a simple indicator to express the level of energy efficiency from the techno-economics perspectives (Miller and Blair, 1985). According to the European Programme ODYSSEE-MURE, Department of Energy and Climate Change (2012), energy efficiency on a technical level is the relationship between the energy consumed and the output produced by that energy. Increasing energy efficiency means using less energy to produce the same level of output.

Analyzing energy efficiency normally requires looking at energy intensity at an economy-wide level, energy consumption per GDP. However, as this can shift due to changes in economic structure, or recession, looking at individual sectors of the economy (energy sector, manufacturing, services, transportation, construction, etc) provides more insights. Moreover, at the level of the aggregate economy, energy efficiency is not a meaningful concept because of the heterogeneous nature of the output. The production of a huge number of goods and the mixing of the transport of freight and people makes an aggregate energy intensity number, a number that disguises rather than illuminates. The energy intensity number based on GDP has little information content without the underlying sector detail. Therefore, efforts attempting to increase efficiency can aim at examining how energy are used in sectoral production (Lenzen, 1998; Gillingham et.al, 2009; Liu et al., 2012).

In an economic system, energy is divided into intermediate demand, that is energy use by sectoral activities to produce goods and services, and final demand that is energy consumed directly by households, the public sector, investment and export. The proportion of intermediate demand out of total energy demand has always been the largest which is shown to be above 50 percent over the last 16 years, as shown in Figure 1.1. As a large fraction of energy is actually directed towards other sectors, it is highly important to study these intermediate demands in the analysis of the sectoral energy efficiencies.

---

[1] Energy produced by sector is for total energy demand that is for intermediate demand and final demand. Intermediate energy demand is the energy demand from sectors to be used as an input to produce their output.

[2] The first and important step taken by the government to mitigate the barrier to efficiency improvement caused by low energy prices is through rationalizing its energy subsidies. The Malaysian government has rationalized its vehicle fuel subsidies since 2010 and removed all subsidies on 1 December 2014. A managed float mechanism has been put in place where prices would adjust according to the market rate. In the meantime, the government is reducing subsidies for natural gas and electricity for the power station and sectors since July 2010.

[3] The growth in total energy demand and GDP is calculated by the author using compounded annual growth rate formula based on the data of total energy demand and GDP, respectively obtained from OECD'd Input-output Table and Bank Negara Annual Report.

[4] Ministry of Energy, Green Technologies and Water: KeTTHA Report – National Energy Efficiency Action Plan (Draft Final Report January 2014), January 2014 Malaysia.

In contrast to the traditional way of analyzing sectoral energy intensities by looking at only direct energy use, this study looks into both direct and indirect energy uses. The indirect energy intensity for a sector represents the inter-sectoral energy relationship. Most people think of energy use in terms of direct purchase of petrol, gasoline, gas and electricity, but a great deal of the sectoral energy use are indirectly embedded in the goods and services that they consume. Cutting down on indirect energy would produce a substantial energy reduction. Analyzing both the direct and indirect energy intensity will provide a holistic assessment of energy use (Liu et al., 2009; Park and Heo, 2007; Pauchari, 2002; Lenzen, 1998; Vringer, 1995).

This study specifically intends to examine the energy intensity performance in terms of total, direct and indirect energy intensity for each sector, and how these energy intensities change over time. The study investigates the energy intensities of 34 sectors between the years 1995 to 2011, employs input-output analysis using input-output tables 1995, 2000, 2005 and 2011 published by the Organisation for Economic Co-operation and Development (OECD).

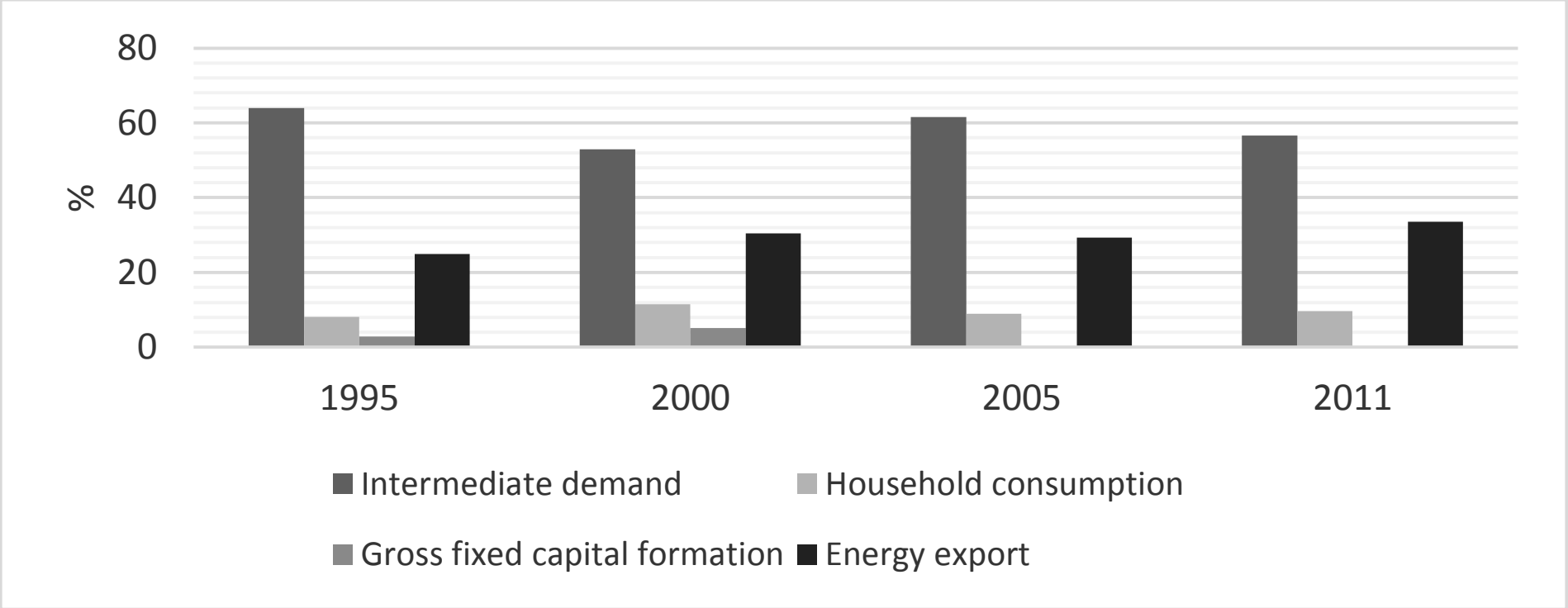

Source: OECD Malaysia Input-output Table, 1995, 2000, 2005, 2011.

Figure 1 Malaysia's Share of Total Energy Demand

## LITERATURE REVIEW

The traditional way of analyzing sectoral energy efficiency only highlights the direct sectoral energy use. However, it is well known that economic systems are highly complex and integrated. Each sector not only consumes energy in a direct way in forms such as electricity, oil, coal, and natural gas but also in an indirect way by consuming energy-intensive intermediate inputs produced by other sectors. Ignoring the linkage between sectors and focusing only on direct energy consumption would undermine the efforts at energy savings (Pauchauri, 2002, Park and Hoe, 2006; Liu et al, 2006, 2012; Lenzen, 1998). According to Bullard and Herendeen (1975) and Skelton (2011), it is crucial to study how different sectors use energy for their operations which should not only focus on the direct energy consumption but also consider the indirect energy consumption so that appropriate national policy can be enacted to promote energy efficient production.

To study the total, direct and indirect energy intensities of sectors, input-output analysis is used in this study. The use of the economic input-output analysis is well established in the energy analysis literature. Early research focused on the determination of energy intensities of goods and services in OECD countries and include those for the US by Wright (1974) and Bullard and Herendeen (1975), and the UK by Pick and Becker (1975).

More recent studies focus on energy consumption pattern represented by direct and indirect energy use with energy efficiency Vringer and Blok (1995) in the Netherlands, Pauchauri and Spreng (2002) in India, Francis (2004) in the UK, Crawford and Treloar (2004) in Australia, Ramirez et al. (2006) in the Netherlands, Park and Hoe (2006) in Korea, and Liu et al. (2009, 2012) in China. Liu et al. (2009) analyze the overall energy intensity of the Chinese alumina refinery plant and found that the overall energy intensity decreased and the energy savings were 49 percent for direct energy and 51 percent for indirect energy. Ramirez et al. (2006) investigate the food and tobacco industry in the Netherlands and found that the cumulative energy savings had been mainly due to improvement efficiency of direct energy use, fossil fuels/heat per unit of product. Liu et al. (2012) show that considerable amount of energy use is embodied in the supply chain especially for construction and other services activities, which is not detected if energy use is allocated on a production basis.

There are two main published studies in which input-output analysis of energy intensity is applied to the Malaysian economy. Moradkhani (2011) calculated the total energy intensities for 54 sector groupings based on the 2000 and 2005 input-output tables, and Norlaila et al. (2012) calculated the total energy and direct energy intensities for 40 sector groupings based on the 1991, 2000 and 2005 input-output tables. The approach discussed in this present study differs from the existing studies in Malaysia in two aspects:

1. In addition to direct energy intensity, this study also includes indirect energy intensity. Moradkhani (2011), and Norlaila et al. (2012) only discussed direct energy use in the form of oil, gas, electricity, coal, and petrol products.
2. This study discussed energy intensity for a longer period from 1995 to 2011, using four input-output tables, for 1995, 2000, 2005, and 2011. Moradkhani used input-output tables for 2000 and 2005 while Norlaila et al. (2012) used input-output tables for 1991, 2000 and 2005.

## METHODOLOGY

This study employs input-output analysis using input-output tables published by the OECD. There are 34 sectors in the Malaysian OECD's Input-output table for the years 1995, 2000, 2005 and 2011. The method of input-output analysis not only measures the direct energy intensity but also identifies all the interactions and the interdependences between sectors in a given economy and hence it allows the estimation of indirect energy intensities of all sectors. Moreover, it helps in the identification of the economic sectors mainly responsible for high energy use in Malaysia.

The energy sector is the totality of all of the sectors involved in the production and sale of energy, including fuel extraction, manufacturing, refining and distribution. We have identified 3 energy sectors; (1) crude petroleum and natural gas, coal, lignite, metal ores, uranium and thorium ores, (2) coke, refined petroleum products (fuel oils, gasoline, illuminating oils, lubricating oils and greases), and nuclear fuel, and (3) electricity and gas. Details of the computational procedures used for calculating the various energy intensities are explained below.

Input-output analysis is an analytical framework for evaluating the interrelations between economic activities, shown by the system of linear equations below:

$$\begin{array}{l} X_1 = x_{11} + x_{12} + x_{13} \dots\dots + x_{134} + Y_1 \\ X_2 = x_{21} + x_{22} + x_{23} \dots\dots + x_{234} + Y_2 \\ X_3 = x_{31} + x_{32} + x_{33} \dots\dots + x_{334} + Y_3 \\ .. \quad .. \quad .. \quad .. \quad .. \quad .. \quad .. \\ .. \quad .. \quad .. \quad .. \quad .. \quad .. \quad .. \\ X_{34} = x_{341} + x_{342} + x_{343} \dots\dots + x_{3434} + Y_{34} \end{array} \tag{1}$$

$X_i$ is the total output of sector *i* and $x_{ij}$ is a sale from sector *i* to sector *j*, $Y_i$ is a sale from sector *i* to final demand.

$$a_{ij} = \frac{x_{ij}}{X_j} \tag{2}$$

$a_{ij}$ is the technical coefficient that is the value of input purchased from all sectors of the economy per unit output of sector *j*. Substituting equation (2) into equation (1) we obtain:

.

$$X_1 = a_{11} X_1 + a_{12} X_2 + a_{13} X_3 \ldots\ldots + a_{134} X_{34} + Y_1$$

$$X_2 = a_{21} X_1 + a_{22} X_2 + a_{23} X_3 \ldots\ldots + a_{234} X_{34} + Y_2$$

$$X_3 = a_{31} X_1 + a_{32} X_2 + a_{33} X_3 \ldots\ldots + a_{334} X_{34} + Y_3$$

.. .. .. .. .. .. .. ..
.. .. .. .. .. .. .. ..

$$X_{34} = a_{341} X_1 + a_{342} X_2 + a_{343} X_3 \ldots\ldots + a_{3434} X_{34} + Y_{34}$$

Transferring all the X's to the left hand-side and re-grouping we obtain:

$$(1-a_{11})X_1 - a_{12}X_2 - a_{13}X_3 \ldots\ldots - a_{134}X_{34} = Y_1$$
$$-a_{21}X_1 + (1-a_{22})X_2 - a_{23}X_3 \ldots\ldots - a_{234}X_{34} = Y_2$$

$$-a_{31}X_1 - a_{32}X_2 + (1-a_{33})X_3 - a_{33}X_3 \ldots\ldots - a_{334}X_{34} = Y_3$$
$$\cdots \quad \cdots\cdot \quad \cdots \quad \cdots \quad \cdots$$

$$-a_{341}X_1 - a_{342}X_2 - a_{343}X_3 \ldots\ldots + (1-a_{3434})X_{34} = Y_{34}$$

In matrix form

$$\begin{bmatrix} (1-a_{11}) - a_{12} - a_{13} \ldots\ldots - a_{134} \\ -a_{21}(1-a_{22}) - a_{23} \ldots\ldots - a_{234} \\ -a_{31} - a_{32}(1-a_{33}) \ldots\ldots - a_{334} \\ \ldots\ldots \\ -a_{341} - a_{342} - a_{343} \ldots\ldots - a_{3433}(1-a_{3434}) \end{bmatrix} \begin{bmatrix} X_1 \\ X_2 \\ X_3 \\ \ldots \\ X_{34} \end{bmatrix} = \begin{bmatrix} Y_1 \\ Y_2 \\ Y_3 \\ \ldots \\ Y_{34} \end{bmatrix} \quad (3)$$

$$(I - A)X = Y$$
$$X = (I - A)^{-1} Y$$
$$X = AX + Y$$

$X$ = vector of output

$Y$ = vector of final demand

The total output of an economy, $X$ can be expressed as the sum of intermediate consumption, $AX$, and final consumption, $Y$.

$A$ is the technical coefficient to represent the direct requirements matrix and $(I - A)^{-1}$ is the interdependence coefficient or also known as Leontief inverse matrix describing the inter-linkages between all sectors of the economy. It shows the relationship between the outputs of the 34 producing sectors and the final demands for the products of these sectors.

The monetary input-output system shown above, equation (3) can be written as in equations (4) and (5),

$$\sum_{j=1}^{3} X_{1j} + \sum_{j=4}^{34} X_{1j} + f_1 - M_1 = X_1 \quad (4)$$

where $X_{1j}$ are the intermediate outputs of sector 1 to be used in the production of goods of sectors 1-34. Meanwhile, $f_1$ is the final demand which includes consumption (household and government), investments and exports for sector 1, and $M_1$ is the import of sector 1. The first summation of Equation (4) means the outputs of the 3 energy sectors, while the second summation means the outputs of the 31 non-energy sectors.

$$\sum_{j=1}^{3} X_{i1} + \sum_{j=4}^{34} X_{i1} + V_1 = X_1 \quad (5)$$

where $X_{i1}$ are the intermediate inputs from sector 1 to 34 for the production of goods of sector 1. $V_1$ is the value-added inputs for the production of goods of sector 1. The first summation of Eq (5) means the inputs of the 3 energy sectors, while the second summation means the inputs of the 31 non-energy sectors.

These conventional monetary input-output tables are then transformed into energy input-output tables with the aid of a uniform (average) energy prices (Miller and Blair, 1985; Park and Heo, 2006; and Liu et al., 2009). The price is calculated using information on energy use by the sectors of the input-output table.

$$P_i = \frac{E_i}{X_{i,} - M_i} \quad (Toe/USD) \quad (6)$$

where $E_i$ is energy use, $P_i$ is the price of energy sector $i$, for example, price of electricity and gas is used to quantify the 34 intermediate inputs of electricity and gas to produce the goods of the 34 sectors. Then the average price is calculated from these prices, which is the uniform price of energy to all 34 sectors ($P_1$ ).
Intermediate energy inputs are then computed using the equation below:

$$\sum_{j=1}^{34} X_{1j} * P_1 = \sum_{j=1}^{34} E_{1j} \quad (7)$$

Once the energy input-output tables are constructed based on equation (7), the direct, indirect and total energy intensities of individual sectors can be calculated. Direct energy intensities are calculated as the ratios of direct energy expenditure converted into the physical term to total inputs.

$$I_1\,(direct) = \sum_{i=1}^{3} \frac{\mathrm{E}_{i,1}}{X_1} \quad (8)$$

Total energy intensities for each energy carrier are calculated with the aid of the Leontief inverse matrices whose elements represents total direct and indirect requirements per USDollar output for each sector.

$$I_i\,(total) = I_i\,(direct) * (1 - A)^{-1} \quad (9)$$

The indirect energy intensities are the difference between total energy intensities and direct energy intensities.

$$I_i\,(indirect) = I_i\,(total) - I_i\,(direct) \quad (10)$$

## FINDINGS

Overall, Malaysia has experienced an increase in total energy intensity from 1995 to 2011 as shown in Figure 4.1. Total energy intensity, expressed in million tonnes of oil equivalent per million US dollar output (Mtoe/USDm) in 2010 constant prices has increased by 10.9 percent from 3.73 Mtoe/USDm in 1995 to 4.13 Mtoe/USDm in 2011. Malaysia's energy intensity of 4.13 Mtoe/USDm was almost double than that of the UK at 2.68 Mtoe/USDm and the US at 2.36 Mtoe/USDm and higher than Indonesia (3.96 Mtoe/USDm), and Thailand (3.11 Mtoe/USDm). The higher energy intensity indicates a higher energy use as a function of its GDP or in other words higher cost of converting energy into GDP as compared to other countries. High energy use can accelerate the negative consequences of energy supply, energy costs and climate change. As mentioned earlier, higher energy used in Malaysia driven mainly by the expansion of manufacturing, transportation and residential sectors (Malaysia National Energy Master Plan, 2014). Moreover, the previous implementation of Malaysian fuel price control policy had accelerated the demand for energy.[5]

The 20 most energy intensive non-energy sectors in Malaysia in terms of their total energy intensity are presented in Table 4.1. The most energy intensive sectors in 2005 and 2011 were transport and storage, other non-metallic mineral products, chemicals and chemical products, financial intermediation, wholesale and retail trade, and agriculture, forestry and fishing. In general, energy intensities increased for many sectors during the period 1995 to 2011 where 17 out of the 20 most intensive non-energy sectors experienced an increased.

Looking at the rankings among the sectors, it can be seen that they did not change much in the period 1995-2011, except for four intensive energy sectors; transport and storage, basic metal, renting of machinery and equipment, and other community, social and personal. Transport and storage was ranked 19th and 8th in 1995 and 2000, respectively but became the most energy intensive in 2005 and 2011. The remarkable increase in transport and storage energy intensity could explain the significant increase in Malaysia's total energy intensity over the period 1995 to 2005. The remarkable shift in transport and storage rank position is mainly due to the significant increase in new passenger and commercial vehicles registered in Malaysia for the year 1995 to 2005, almost double increase from 285,792 units to 552,316 units (Malaysia Automative Association, 2018). The total energy intensity of renting of machinery and equipment, and other community, social and personal, , increased by about six and four times higher, respectively changing their ranks from 27th to 13th, and 21st to 9th between 1995 and 2011. Nevertheless, an opposite trend is shown by basic metal total energy intensity where there was a huge decline from 240.3 ktoe/USDm in 1995 to 68.8 ktoe/USDm in 2011, changing its rank from 1st to 16th. The continued decrease in Malaysia's steel production led the decrease in basic metal energy intensity. Malaysian steel mills had trim their production and closed down production lines due to low capacity utilization as a result of intense competition from cheaper and below-cost steel imports, mainly from China (Department of Statistics, 2012).

The increase in total energy intensity of many sectors contributed to the increase in Malaysia's total energy intensity, indicating a reduction in the energy efficiency of the country. If we consider that the increase in Malaysia's total energy intensity was moderated by the huge reduction in basic metal energy intensity, and the reduction in energy intensity of basic metal sector was due to lower production and not due to an increase in efficiency, we could say that the reduction in energy efficiency of the country was even bigger.

Out of Malaysia's 20 most energy intensive non-energy sectors, 12 sectors recorded a higher share of direct energy intensity to total energy intensity for most years as shown in Table 4.2. The share was high for the three most intensive energy intensity sectors; transport and storage, other non-metallic mineral products, and chemicals and chemical products, and the shares were still high although they declined for the period 1995 to 2011. Other sectors where direct energy intensity share was also high were other community, social and personal services, construction, renting of machinery and equipment, hotel and restaurants, basic metals, textiles, textile products, leather and footwear, and other manufacturing.

Certain sectors gain importance upon considering the indirect energy use. For instance, in 2011, the financial intermediation sector was the 17th most important sector in terms of the direct energy intensity but it was even more significant at the 4th position in terms of the total energy intensity. In other words, this shows that the indirect energy intensity provides significant contribution towards the high total energy intensity of the financial intermediation sector. This means that the development of the financial intermediation sector has increased the opportunities for investment and has supported increased lending to households and firms, thus, encouraging consumers to purchase expensive or major items such as automobiles and machinery, and thereby increasing energy consumption. The study by Chang (2015) also reveals that energy consumption increases with financial development in the non–high income regime.

Other sectors that recorded low share of direct energy intensity or in other words have high share of indirect energy intensity are food products, beverage and tobacco (8.4 percent), real estate activities (21 percent), fabricated metal products (27 percent), post and telecommunications (27.3 percent) and wholesale and retail products (33.7 percent). These sectors are also among the intensive energy sectors as observed from their total energy intensities by ranking (Table 4.1). It is interesting to note that the total energy intensity rank of the transport and storage sector moves up substantially due to the increase in its indirect energy intensity. Similarly, other non-metallic mineral products sector, switched its total energy intensity rank to a higher position from 5th in 1995 to 2nd in 2011, based on an increase in indirect intensity (Table 4.1 and Table 4.2). This indicates the increasing importance of indirect energy intensity to these two most intensive energy sectors.

[5] The first and important step taken by the government to mitigate the barrier to efficiency improvement caused by low energy prices is through rationalizing its energy subsidies. The Malaysian government has rationalized its vehicle fuel subsidies since 2010 and removed all subsidies on 1 December 2014. A managed float mechanism has been put in place where prices would adjust according to the market rate. In the meantime, the government is reducing subsidies for natural gas and electricity for the power station and sectors since July 2010.

Table 1 Malaysia's sectoral total energy intensity of most intensive non-energy sectors

| Sector | 1995 | | 2000 | | 2005 | | 2011 | |
|---|---|---|---|---|---|---|---|---|
| | ktoe/USDm | ranks | ktoe/USDm | ranks | ktoe/USDm | ranks | ktoe/USDm | ranks |
| Transport and storage | 25.4 | 19 | 105.9 | 8 | 336.1 | 1 | 339.9 | 1 |
| Other non-metallic mineral products | 132.5 | 5 | 261.4 | 2 | 279.4 | 2 | 261.7 | 2 |
| Chemicals and chemical products | 231.2 | 2 | 156.0 | 5 | 250.2 | 3 | 243.3 | 3 |
| Financial intermediation | 179.2 | 3 | 186.1 | 3 | 163.2 | 4 | 161.7 | 4 |
| Agriculture, hunting, forestry and fishing | 126.2 | 6 | 263.9 | 1 | 149.6 | 5 | 154.7 | 5 |
| Wholesale and retail trade; repairs | 135.3 | 4 | 74.9 | 15 | 100.6 | 11 | 148.7 | 6 |
| Fabricated metal products | 56.8 | 8 | 96.0 | 11 | 115.9 | 6 | 131.0 | 7 |
| Food products, beverages and tobacco | 41.9 | 12 | 149.9 | 6 | 97.8 | 13 | 109.7 | 8 |
| Other community, social and personal services | 23.1 | 21 | 52.9 | 20 | 100.8 | 10 | 103.1 | 9 |
| Construction | 54.5 | 10 | 98.0 | 10 | 114.9 | 7 | 97.3 | 10 |
| R&D and other business activities | 25.7 | 18 | 85.3 | 12 | 92.5 | 14 | 94.3 | 11 |
| Rubber and plastics products | 34.9 | 15 | 106.4 | 7 | 82.6 | 16 | 90.8 | 12 |
| Renting of machinery and equipment | 12.7 | 27 | 12.3 | 30 | 82.8 | 15 | 79.4 | 13 |
| Hotels and restaurants | 63.7 | 7 | 75.3 | 14 | 106.2 | 9 | 78.9 | 14 |
| Post and telecommunications | 47.8 | 11 | 62.5 | 18 | 74.4 | 17 | 76.5 | 15 |
| Basic metals | 240.3 | 1 | 163.8 | 4 | 99.2 | 12 | 68.8 | 16 |
| Textiles, textile products, leather and footwear | 20.3 | 23 | 80.1 | 13 | 65.5 | 18 | 56.1 | 17 |
| Manufacturing nec; recycling | 17.1 | 26 | 37.4 | 24 | 108.7 | 8 | 49.3 | 18 |
| Pulp, paper, paper products, printing and publishing | 36.8 | 14 | 101.5 | 9 | 45.2 | 22 | 48.1 | 19 |
| Real estate activities | 20.8 | 22 | 64.5 | 17 | 54.4 | 19 | 46.2 | 20 |

Source: own calculation based on Input-output data OECD, 1995, 2000, 2005, 2011.

Table 2 Malaysia's sectoral direct energy intensity (ktoe/USDm) and ratios of direct energy intensity to the total energy intensity (%)

| Sector | 1995 | | 2000 | | 2005 | | 2011 | |
|---|---|---|---|---|---|---|---|---|
| | ktoe/USDm | % | ktoe/USDm | % | ktoe/USDm | % | ktoe/USDm | % |
| Transport and storage | 22.5 | 88.6 | 73.8 | 69.7 | 203.8 | 60.6 | 232.2 | 68.3 |
| Other non-metallic mineral products | 118.3 | 89.3 | 211.3 | 80.8 | 223.2 | 79.9 | 195.2 | 74.6 |
| Chemicals and chemical products | 194.0 | 83.9 | 98.4 | 63.0 | 138.1 | 55.2 | 136.4 | 56.1 |
| Financial intermediation | 19.7 | 11.0 | 26.9 | 14.4 | 25.3 | 15.5 | 23.0 | 14.2 |
| Agriculture, hunting, forestry and fishing | 22.5 | 17.8 | 84.6 | 32.1 | 74.1 | 49.5 | 64.2 | 41.5 |
| Wholesale and retail trade; repairs | 18.2 | 13.5 | 37.9 | 50.6 | 46.0 | 45.8 | 50.2 | 33.7 |
| Fabricated metal products | 42.5 | 74.8 | 59.4 | 61.9 | 42.6 | 36.8 | 35.4 | 27.0 |
| Food products, beverages and tobacco | 29.6 | 70.7 | 51.5 | 34.4 | 18.8 | 19.2 | 9.2 | 8.4 |
| Other community, social and personal services | 13.4 | 58.3 | 44.0 | 83.2 | 61.1 | 60.6 | 69.0 | 66.9 |
| Construction | 20.9 | 38.4 | 62.4 | 63.6 | 75.6 | 65.9 | 69.7 | 71.6 |
| R&D and other business activities | 17.3 | 67.3 | 33.9 | 39.8 | 37.1 | 40.1 | 39.5 | 42.0 |
| Rubber and plastics products | 17.1 | 49.0 | 73.8 | 69.4 | 40.5 | 49.0 | 33.5 | 36.9 |
| Renting of machinery and equipment | 7.2 | 56.7 | 6.7 | 54.5 | 80.2 | 96.9 | 77.2 | 97.3 |
| Hotels and restaurants | 52.7 | 82.8 | 50.0 | 66.4 | 89.3 | 84.0 | 54.4 | 68.9 |
| Post and telecommunications | 38.5 | 80.4 | 27.0 | 43.3 | 19.8 | 26.7 | 20.9 | 27.3 |
| Basic metals | 221.6 | 92.2 | 103.0 | 62.9 | 72.1 | 72.7 | 61.5 | 89.5 |
| Textiles, textile products, leather and footwear | 10.3 | 50.6 | 57.2 | 71.4 | 38.1 | 58.2 | 41.7 | 74.4 |
| Manufacturing nec; recycling | 12.4 | 72.1 | 27.2 | 72.8 | 84.6 | 77.8 | 42.5 | 86.2 |
| Pulp, paper, paper products, printing and publishing | 11.7 | 31.9 | 57.5 | 56.6 | 38.1 | 84.3 | 34.8 | 72.2 |
| Real estate activities | 11.0 | 53.0 | 38.3 | 59.4 | 9.2 | 16.9 | 9.7 | 21.0 |

Source: own calculation based on Input-output data OECD, 1995, 2000, 2005, 2011

## CONCLUSION

This study analyzes the total, direct and indirect energy requirements of sectors to understand the structure and evolution of sectoral energy use in Malaysia. As mentioned in the study, the amount of direct energy use per unit of output incorrectly measures the energy intensity of an activity. Discounting the indirect energy use underestimates the energy intensity. Hence, measurement of energy intensity based on total energy use, including the indirect energy use, is an appropriate measure of energy intensity.

Our findings show that in general, total, direct, and indirect energy intensities of sectors show an increasing trend. Looking at the trend of intensities in detail show that they have increased substantially from 1995 to 2005, reflecting the significantly lower value-added production structure in 1995 to 2005 as the value added of Malaysian production grew much slower than the energy input used. A lower value-added production structure that is shown by the increasing trend in sectoral energy intensity means that the sector requires more energy to produce a dollar of value-added in the production process. Although Malaysia moved to a slightly higher value added production structure in 2005 to 2011, its total energy intensity was still high, and even higher than the 1995 level.

It is the share of sectoral direct energy use that determines the fluctuations in the total sectoral energy intensity for most sectors for all years. For instance, for the three most energy intensive non-energy sectors; transport and storage, other non-metallic mineral products, and chemicals and chemical products, the contribution of direct energy use to the total energy use was high, mostly more than 60 percent for most years. This mirrors the importance of determining total energy intensities as lowering sectoral direct energy use could be the most effective strategy for reducing total energy use. Nevertheless, there were also energy intensive sectors that showed higher indirect energy use as compared to direct energy use, particularly, financial intermediation, agriculture, hunting and forestry, wholesale and retail trade, and food, beverage, and tobacco. Therefore, the importance of lowering indirect energy use cannot simply be neglected. Moreover, the two most intensive energy sectors, transport and storage, and other non-metallic mineral products indicate an increased importance of indirect energy intensity in their total energy intensities.

The results provide an innovative perspective for explaining the rapid growth of ‘‘energy- intensive’’ sectors in Malaysia. While transportation and heavy sectors, such as transportation and storage, basic metals, and other non-metallic mineral products, are typical energy-intensive sectors and have higher level of total and direct energy use, sectors such as financial intermediation, agriculture, hunting and forestry, wholesale and retail trade, and food, beverage, and tobacco have relatively higher total and indirect energy use and should be considered energy-intensive sectors from a supply chain perspective. Therefore, our results indicate that some sectors that are not traditionally regarded as ‘‘energy-intensive’’ are actually sectors with higher embodied energy consumption, and, hence, these sectors should be regarded as ‘energy-intensive’. Besides that, the results may also be translated as some sectors that are traditionally regarded as ‘clean-sector’ are actually ‘dirty-sector’. Therefore, energy efficiency policies should be revised based on such energy intensities (both direct and indirect energy intensity), so that rational consideration can be given to these sectors. This study can be used by decision-makers to reconsider their energy management targets and regulations by considering such sectoral energy intensities. From this, we could say that our results provide a more holistic picture of sectoral energy use and therefore can help decision-makers in designing more appropriate policies.